**TITLE**
Controlling transition metal atomic ordering in two-dimensional $Mo_{1-x}W_xS_2$ alloys

**AUTHORS**
Jeff J.P.M. Schulpen[1], Marcel A. Verheijen[1,2], Wilhelmus M.M. Kessels[1], Vincent Vandalon[1], Ageeth A. Bol[1]

**AFFILIATION**
[1] Department of Applied Physics, Eindhoven University of Technology, Eindhoven, The Netherlands.
[2] Eurofins Materials Science BV, High Tech Campus, Eindhoven, The Netherlands

**ABSTRACT**
The unique optical and electronic properties of two-dimensional transition metal dichalcogenides (2D TMDs) make them promising materials for applications in (opto-)electronics, catalysis and more. Specifically, alloys of 2D TMDs have broad potential applications owing to their composition-controlled properties. Several important challenges remain regarding controllable and scalable fabrication of these alloys, such as achieving control over their atomic ordering (i.e. clustering or random mixing of the transition metal atoms within the 2D layers). In this work, atomic layer deposition (ALD) is used to synthesize the TMD alloy $Mo_{1-x}W_xS_2$ with excellent composition control along the complete composition range $0 \leq x \leq 1$. Importantly, this composition control allows us to control the atomic ordering of the alloy from well-mixed to clustered while keeping the alloy composition fixed, as is confirmed directly through atomic-resolution HAADF-STEM imaging. The control over atomic ordering leads to tuning of the bandgap, as is demonstrated using optical transmission spectroscopy. The relation between this tuning of the electronic structure and the atomic ordering of the alloy was further confirmed through ab-initio calculations. Furthermore, as the atomic ordering modulates from clustered to well-mixed, the typical $MoS_2$ and $WS_2$ $A_{1g}$ vibrational modes converge. Our results demonstrate that atomic ordering is an important parameter that can be tuned experimentally to finely tune the fundamental properties of 2D TMD alloys for specific applications.

**INTRODUCTION**
Two-dimensional (2D) materials consist of molecular layers whose mutual interaction is weak compared to the bonding within the layers,[1] such that these materials can be thinned down to a single molecular layer while maintaining predictable properties[2] which often include interesting physical phenomena induced by quantum confinement.[3,4] Transition metal dichalcogenides (TMDs) are a particularly important group of 2D materials since many of them, including the well-studied materials $MoS_2$ and $WS_2$, are semiconductors. This makes them interesting for a wide array of applications such as ultimately-scaled nanoelectronics,[5] versatile nanophotonics,[6] and highly efficient photovoltaics.[7]

Alloying of semiconducting 2D TMDs further expands the possible applications of these materials.[29] For example, their electronic bandgap can be tuned continuously over a wide energy range both through chalcogen substitution[14,16-22,61] and through metal substitution.[23-25] Furthermore, alloying can improve the quality of TMD films, such as by reducing the density of chalcogen vacancy defects.[15]

Important challenges remain in achieving controllable and scalable synthesis of TMD alloys. Techniques such as chemical vapor deposition (CVD) and chemical vapor transport (CVT), which are known to produce high-quality samples of pure TMDs[8-10], are being used to synthesize such alloys[26-28]. However, the control over alloy composition is usually limited when using these techniques. Furthermore, it has not been possible to control the atomic ordering of the alloys (i.e. well-mixed or clustered) using these techniques.[29] Instead, the alloys grown by CVD and CVT invariably exhibit random mixing of the constituents, likely due to strong surface diffusion of atoms during synthesis owing to the inherently high processing temperatures of these techniques.[30-32]

Atomic layer deposition (ALD) is a thin film deposition technique that exploits self-limiting surface reactions of vapor-phase precursors to achieve sub-nanometer growth control at low deposition temperatures (< 450 °C). While direct ALD growth of pure TMDs has been reported,[11-12,34] efforts on ALD of TMD alloys have focused on high-temperature sulfurization of metal (oxide) parent films, with the parent film synthesized by ALD.[33] This approach yields a higher level of control than is achieved with CVD, for example by allowing a graded composition profile of the alloy.[33] Nevertheless, the high-temperature sulfurization step diminishes the low-temperature benefit of ALD, such that there remains a need for a fully low-temperature ALD synthesis process for better control over the composition and atomic ordering of the TMD alloys.

In this work, we introduce a plasma-enhanced atomic layer deposition (PE-ALD) process for one-step deposition of $Mo_{1-x}W_xS_2$ alloys at a low substrate temperature of 350 °C without the need for post-deposition sulfurization. Using this process, we demonstrate fine control over the alloy composition $x$ by manipulating the relative number of $MoS_2$ and $WS_2$ deposition cycles in a supercycle scheme, which is an established method of growing alloys of e.g. oxides by ALD.[14] Furthermore, we demonstrate for the first time control over the atomic ordering within the $Mo_{1-x}W_xS_2$ layers, which is confirmed directly through high-angle annular dark-field scanning transmission electron micrography (HAADF-STEM). The control over the atomic ordering of the alloys is achieved by manipulating the supercycle period of the ALD process (i.e. the total number of ALD cycles per supercycle). Beyond HAADF-STEM imaging, we report the effects of atomic ordering on the electronic structure and phonon spectrum of the alloy, which are probed through optical transmission spectroscopy and Raman spectroscopy respectively. Finally, density functional theory (DFT) calculations are carried out to provide fundamental insight into how the electronic structure of the alloys is modulated by their atomic ordering.

**RESULTS**
$Mo_{1-x}W_xS_2$ alloys were grown by ALD using a scheme whereby supercycles consisting of $n$ $MoS_2$ ALD cycles followed by $m$ $WS_2$ ALD cycles are repeated, as illustrated in figure 1. The supercycle fraction $n/(n+m)$ determines the composition of the alloy (the Mo/W ratio), whereas the supercycle length $n+m$ will be shown to control the atomic ordering of the alloy. The total number of ALD cycles $N \times (n+m)$ determines the thickness of the deposited film.

**I. CONTROL OVER ALLOY COMPOSITION**
A series of six $Mo_{1-x}W_xS_2$ samples was prepared using supercycle fractions of 0.0 (pure $WS_2$), 0.2, 0.4, 0.6, 0.8 and 1.0 (pure $MoS_2$), each for a total of 50 ALD cycles. We first investigate the growth and chemical composition of the alloys, after which we characterize their composition-dependent vibrational spectrum and electronic structure.

**In-situ growth characterization**
The growth of the $Mo_{1-x}W_xS_2$ films by ALD was characterized in-situ by spectroscopic ellipsometry (SE). Figure 2a shows the deposited film thickness per ALD cycle (GPC) as a function of the cycle fraction $n/(n+m)$. The GPC of the pure $WS_2$ and $MoS_2$ and films are determined to be 0.08 nm and 0.13 nm respectively, in line with our previously reported characterization of these ALD processes.[11,34] The GPC of the alloy increases linearly with the molybdenum content, which is in line with the rule of mixtures for alloys:[42]

$$GPC_{alloy} = \frac{n}{n+m} \times GPC_{MoS_2} + \frac{m}{n+m} \times GPC_{WS_2}, \quad (1)$$

demonstrating well-behaved alloying behavior free from the non-idealities commonly observed when depositing alloys by ALD.[42]

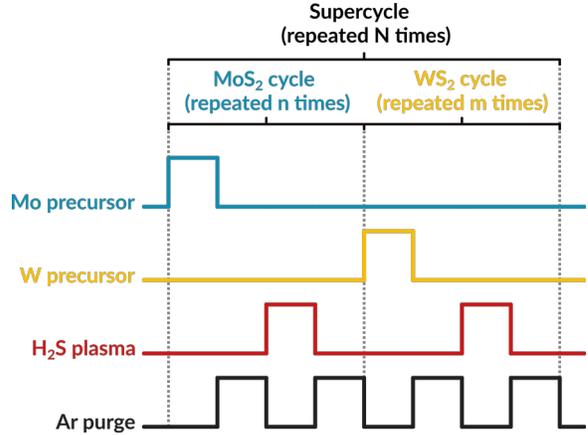

Figure 1: The ALD supercycle scheme used for growth of $Mo_{1-x}W_xS_2$ alloys by alternating ALD cycles of $MoS_2$ and $WS_2$. The cycle fraction n/(n+m) determines the composition x of the alloy, whereas the number of cycles per supercycle n+m (i.e. the supercycle length) is exploited to control the atomic ordering of the alloy. The total number of cycles N×(n+m) determines the film thickness.

**Chemical composition**
On the same set of samples, the alloy composition $x$ and stoichiometry (defined as the S/(Mo+W) ratio) as a function of the supercycle fraction $n/(n+m)$ were measured using x-ray photoelectron spectroscopy (XPS) and are shown in figure 2b. The trend in alloy composition x closely follows the rule of mixtures given by

$$x = \frac{[Mo]}{[Mo]+[W]} = \frac{\frac{n}{n+m} \times GPC_{at,Mo}}{\frac{n}{n+m} \times GPC_{at,Mo} + \frac{m}{n+m} \times GPC_{at,W}}, \quad (2)$$

where the $GPC_{at,Mo}$ is the number of Mo atoms deposited per $MoS_2$ ALD cycle, and $GPC_{at,W}$ the number of W atoms deposited per $WS_2$ ALD cycle. By fitting equation (2) to the data, the atomic GPC ratio between Mo and W is found to be 1.67 ± 0.03. This value is within experimental error of the GPC ratio found from the ellipsometry data, indicating that the trend in film thickness as a function of the cycle fraction can be attributed to the difference between the number of Mo and W atoms deposited per ALD cycle. As also shown in figure 2b, the deposited alloys are approximately stoichiometric (sulfur to metal ratio ≈ 2), with a trend towards over-stoichiometry (excess sulfur) for tungsten-rich films. Analogously to equation (1), a rule of mixtures can be formulated for the stoichiometry:

$$s_{alloy} = \frac{n}{n+m} \times s_{MoS_2} + \frac{m}{n+m} \times s_{WS_2}, \quad (3)$$

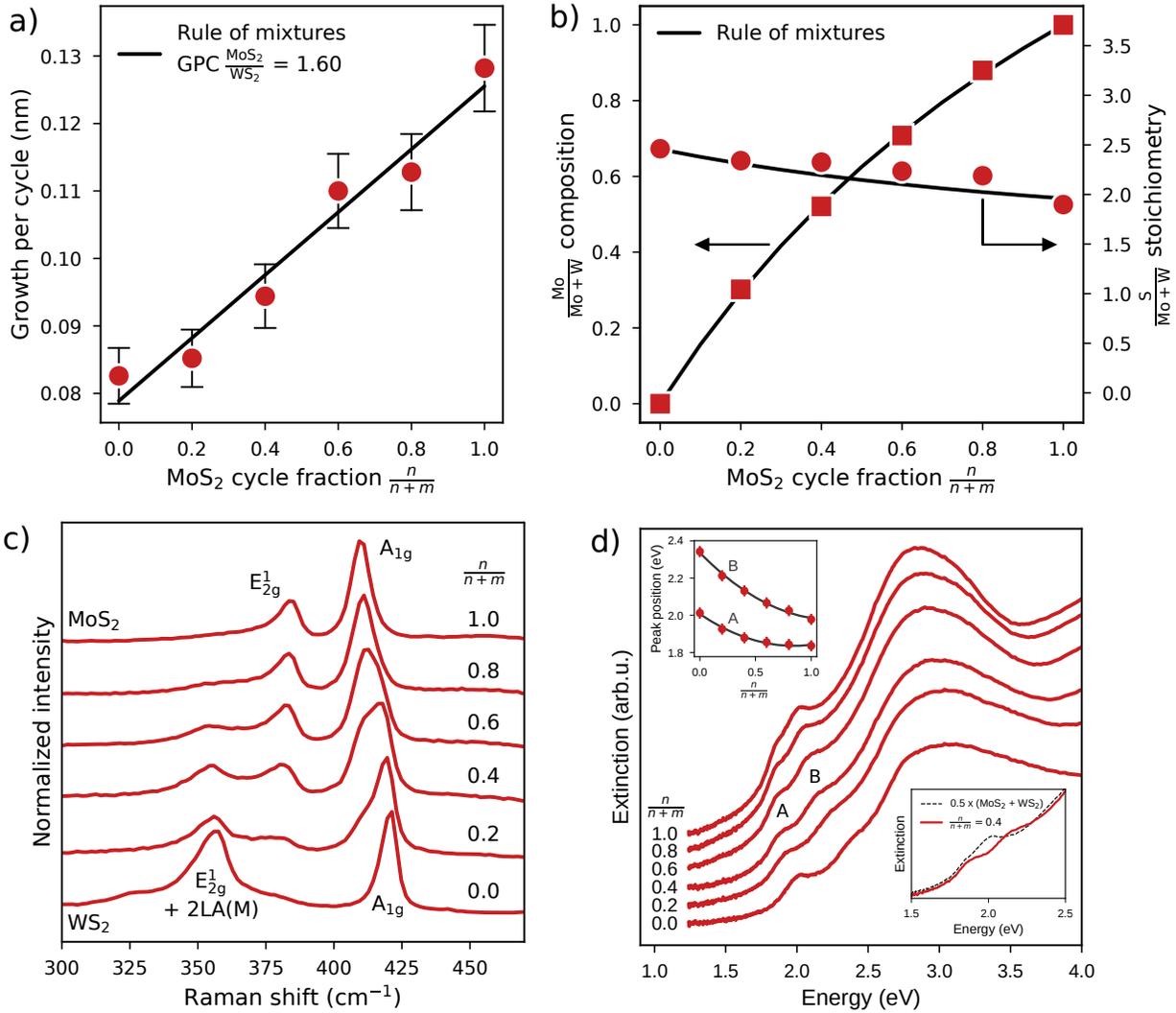

Figure 2: Impact of the composition x (controlled through the ALD cycle fraction n/(n+m)) of the $Mo_{1-x}W_xS_2$ films on their a) growth per cycle measured by in-situ spectroscopic ellipsometry, b) elemental composition measured by x-ray photoelectron spectroscopy, c) vibrational spectrum measured by Raman spectroscopy, d) electronic structure measured by optical extinction spectroscopy. The insets in d) show the positions of the absorption peaks as a function of alloy composition, as well as the difference between the alloy spectrum and a superposed $MoS_2 + WS_2$ spectrum. Spectra in c) and d) have been offset vertically for legibility.

where $s_{MoS2}$ and $s_{WS2}$ are the stoichiometries of the pure $MoS_2$ and $WS_2$ films respectively. The stoichiometry of the alloys as a function of the cycle fraction is well described by equation (3), again confirming the well-behaved growth of the $Mo_{1-x}W_xS_2$ alloys over the complete composition range $0 \leq x \leq 1$.

**Crystallinity and vibrational spectrum**

The impact of the alloy composition on the crystallinity and vibrational spectrum of $Mo_{1-x}W_xS_2$ alloys was studied by Raman spectroscopy: the spectra are shown in figure 2c. In the spectrum of the pure $MoS_2$ film, two dominant peaks at 384 cm$^{-1}$ and 409 cm$^{-1}$ are observed which can be identified as the in-plane $E^1_{2g}$ mode and out-of-plane $A_{1g}$ mode respectively.[43] Peaks corresponding to the same vibrational modes are observed in the spectrum of the pure $WS_2$ films at 356 cm$^{-1}$ and 420 cm$^{-1}$.[44] It should be noted that the $E^1_{2g}$ peak of $WS_2$ overlaps with its 2LA(M) mode, which dominates when using 514 nm excitation light on pure $WS_2$.[45,46] For simplicity, we nevertheless refer to the peak complex at 356 cm$^{-1}$ as the $E^1_{2g}$ peak. The $E^1_{2g}$ and $A_{1g}$ peaks of $MoS_2$ and $WS_2$ persist at all alloy fractions $0 < x < 1$, indicating a polycrystalline structure of the alloys along the complete composition range.

As a function of the alloy fraction $x$, the MoS$_2$-like and WS$_2$-like $E^1_{2g}$ peaks remain spectrally separated and their relative intensities scale with the alloy fraction, i.e. they exhibit two-mode behavior.[47] Similarly, though the spectral proximity of the A$_{1g}$ modes makes their deconvolution more challenging, the asymmetric lineshape of the A$_{1g}$ complex at intermediate alloy fractions suggests two-mode behavior for this mode as well. The same behavior is observed from alloys deposited by CVT.[28]

**Electronic structure**
To study the evolution of the electronic structure of the alloys as a function of their composition, optical transmission measurements were carried out: the spectra are shown figure 2d. In the spectrum of pure MoS$_2$, two extinction peaks are seen around 1.86 eV and 2.02 eV, which can be identified as the fundamental electronic transitions of the A-exciton and B-exciton.[57-58,63] Analogous peaks are observed in the spectrum of pure WS$_2$ around 2.01 eV and 2.38 eV, which is in line with literature values for the A and B exciton energies of this material.[57-58,63]

The spectra of intermediate alloys each exhibit two distinct extinction peaks in the spectral region between 1.8 eV and 2.5 eV, analogous to the A and B excitonic peaks of pure MoS$_2$ and WS$_2$. Furthermore, the positions of these peaks vary smoothly with the alloy composition. This behavior signifies good mixing of the alloy's constituents, since any significant phase separation in the alloy would result in a spectrum that resembles a superposition of the spectra of the constituents, which is not observed. Additionally, the shifts of the excitonic peaks as a function of alloy composition follow a quadratic bowing trend, in line with theoretical calculations.[25]

**II. Control over atomic ordering**
The ideal rule-of-mixtures alloying behavior of the described deposition process opens up interesting new possibilities of fine-tuning the growth of the alloy. As demonstrated in the previous section, the cycle fraction of the supercycle process controls the composition of the alloy. On the other hand, it is known that the supercycle period can be used to control the ordering of an alloy: short supercycle periods produce a well-mixed film while long supercycle periods produce a nanolaminate film.[42] Usually,[42] non-ideal growth characteristics such as heteronucleation delays lead to changes in composition when the supercycle length is changed, such that the effects of the cycle fraction and the supercycle length are not independent. Since our deposition process for Mo$_{1-x}$W$_x$S$_2$ shows no such non-ideal behavior, it is a very interesting case study into the isolated effects of supercycle length at a fixed alloy composition. Furthermore, since it takes approximately 15 ALD cycles to grow a closed monolayer film with our process, a short supercycle (< 10 cycles) process could be used to tune the atomic ordering within the individual molecular layers of the alloy without changing their composition. To confirm that the supercycle length only influences the atomic ordering of the Mo$_{1-x}$W$_x$S$_2$ alloys and not their composition, XPS and Rutherford backscattering spectroscopy (RBS) measurements were carried out. These measurements indicate that the composition $x$ changes by at most a few percent between supercycle lengths of 2 and 48 cycles (see also the Supplementary Information).

**HAADF-STEM imaging**
To directly study the atomic ordering of the ALD-grown Mo$_{1-x}$W$_x$S$_2$ films, HAADF-STEM imaging was performed on two monolayer samples grown using different supercycle lengths. The first was made through a single supercycle of 10 ALD cycles, i.e. 5 MoS$_2$ cycles followed by 5 WS$_2$ cycles. The second was made through 5 supercycles of 2 cycles each, i.e. 5 x (1 MoS$_2$ cycle + 1 WS$_2$ cycle), again for a total of 10 ALD cycles. In the atomic-resolution STEM micrographs (figure 6a-b), Mo and W atoms can be distinguished by their Z-contrast: the heavier W atoms appear brighter than the lighter Mo atoms, while S atoms are not visible in HAADF-STEM mode.

The hexagonal crystal structure of MoS$_2$ and WS$_2$ is clearly visible in the STEM micrographs, directly confirming the polycrystallinity of the deposited alloys: the crystal (grain) sizes are approximately 5 to 10 nm. Comparing the STEM images of the long-supercycle sample (figure 6a) and the short-supercycle sample (figure 6b) shows a clear difference in atomic ordering of the alloys. The long-supercycle sample exhibits clusters of Mo atoms which are bordered by W atoms. The short-supercycle sample does not exhibit such clustering of Mo and W atoms: instead the Mo and W atoms appear to be randomly distributed in the alloy.

Considering that the long-supercycle sample was grown by performing 5 MoS$_2$ cycles followed by 5 WS$_2$ cycles, the distribution of Mo and W in this sample is consistent with MoS$_2$ island growth during the first 5 cycles followed by epitaxial growth of WS$_2$ at the edges of the MoS$_2$ islands in the last 5 ALD cycles. The result is the formation of 2D core/shell-like nanoparticles with MoS$_2$ cores and WS$_2$ shells. Conversely, for the short-supercycle sample, the number of consecutive MoS$_2$ and WS$_2$ cycles is too low for significant clusters of MoS$_2$ or WS$_2$ to form. To support this proposed mechanism, simulations of the initial film growth were conducted using a simplified two-dimensional growth model based on sequential precursor adsorption. The results of these simulations are shown in figure 6e-f. Indeed, the observed differences in atomic ordering as a function of supercycle length are reproduced by this model. In summary, the atomic ordering within the individual $Mo_{1-x}W_xS_2$ molecular layers can be controlled through the supercycle length $n+m$ of the ALD process.

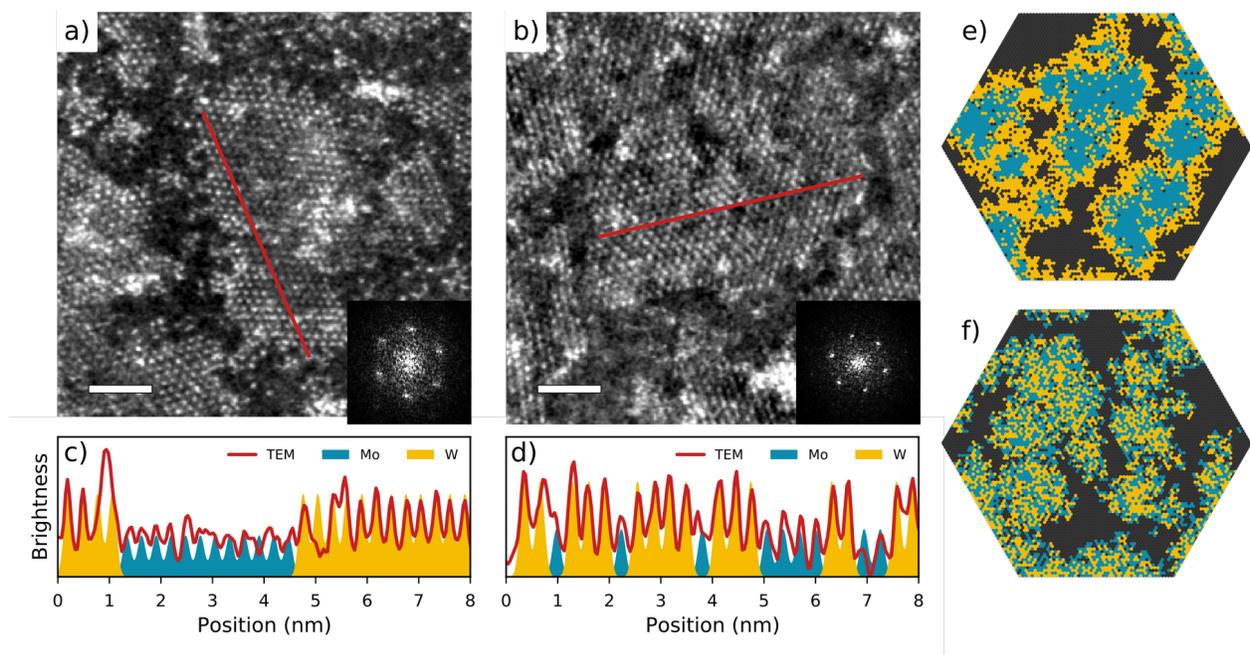

*Figure 3: HAADF-STEM micrographs of monolayer $Mo_{1-x}W_xS_2$ deposited using supercycle lengths of 10 cycles (a) and 2 cycles (b), both for a total of 10 ALD cycles. Scalebars are 2 nm. The insets show local Fourier transforms, demonstrating the hexagonal crystal structure of both samples. The longer supercycle length results in clustering of Mo and W while the shorter supercycle length produces a well-mixed alloy, as also illustrated by the intensity profiles (c,d). Results of stochastic simulations of the film growth (e,f) reproduce the observed differences in atomic ordering. Additional HAADF-STEM micrographs are provided in the Supplementary Information.*

**Impact of atomic ordering on the vibrational structure**

The influence of the atomic ordering of the $Mo_{1-x}W_xS_2$ alloys on their vibrational structure was investigated by Raman spectroscopy. Few-layer (6-8L) samples were grown using supercycle lengths of 2, 6, 10, 16 and 24 cycles at a constant $MoS_2$ cycle fraction $n/(n+m)$ of 0.5 and a total number of 48 ALD cycles for each sample. The Raman spectra are shown in figure 4. For the sample with the longest supercycle length of 24 cycles, four peaks are seen at frequencies corresponding to the $E^1_{2g}$ and $A_{1g}$ modes of bulk $MoS_2$ and $WS_2$,[44,49] indicating that the sample resembles a heterostructure as is expected for long supercycles.[56] As the supercycle period decreases, the $MoS_2$-like $E^1_{2g}$ vibration at 383 cm$^{-1}$ redshifts by 3.0 ± 0.5 cm$^{-1}$ along with a broadening of the peak. The $WS_2$-like $E^1_{2g}$ peak does not shift as much as the $MoS_2$-like $E^1_{2g}$ peak, though some redshift is still observed.

The $A_{1g}$ peaks are spectrally separated at a supercycle period of 24 cycles, while for shorter supercycle periods the two individual peaks cannot be resolved. This merging of the two $A_{1g}$ phonon peaks cannot be caused purely by a broadening of these peaks, since that would broaden the total two-peak complex, which is not observed. Hence, a frequency shift of one or both of the $A_{1g}$ peaks must be involved. Spectral deconvolution indicates the $MoS_2$-like $A_{1g}$ mode blueshifts by 3.5 ± 0.5 cm$^{-1}$ while the $WS_2$-like $A_{1g}$ mode redshifts by 2.0 ± 0.3 cm$^{-1}$ as the supercycle length is shortened from 24 to 2 cycles.

The broadening of the two $E^1_{2g}$ peaks for short supercycle lengths can be understood as a consequence of the enhanced mixing of the alloy constituents, which reduces the crystalline order and relaxes the momentum selection rule on the Raman scattering process.[50] On the other hand, the redshift of the $E^1_{2g}$ peaks as a function of the supercycle period cannot easily be explained on the basis of known effects. Firstly, strain[51–53] of the alloy film is not expected to be the cause of these shifts since $MoS_2$ and $WS_2$ have identical crystal structures with in-plane lattice constants differing by only 0.2%.[54] Furthermore, opposite strain on the constituents $MoS_2$ and $WS_2$ would lead to opposite shifts in the two $E^1_{2g}$ peaks, but parallel shifts are observed. Secondly, an increase in sulfur vacancy concentration[55] can be ruled out as the cause of the observed $E^1_{2g}$ and $A_{1g}$ shifts since the stoichiometry was found to remain constant at 2.0 ± 0.1 between supercycle periods of 2 and 12 cycles from Rutherford backscattering (RBS) analysis. Additionally, an increase in sulfur vacancy concentration would shift the $E^1_{2g}$ frequencies more than the $A_{1g}$ frequencies,[55] which we do not observe. Having ruled out these alternative causes, we conclude that the $E^1_{2g}$ and $A_{1g}$ Raman peaks of $Mo_{1-x}W_xS_2$ are sensitive to changes in atomic ordering of the alloy, making Raman spectroscopy an accessible tool for probing the atomic ordering of 2D alloys. Additionally, these findings indicate that the control over atomic ordering which was demonstrated for monolayer films (figure 3) extends to the few layer (6-8L) regime.

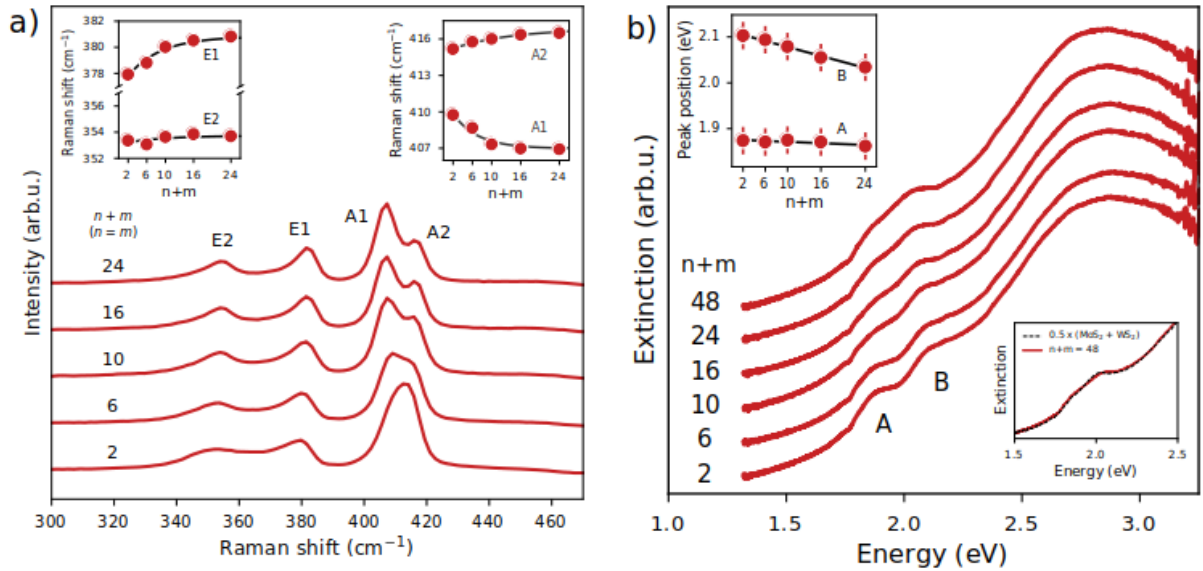

*Figure 4: Effect of atomic ordering (i.e. supercycle length) on a) the vibrational spectrum (Raman spectroscopy) and b) the electronic structure (optical extinction spectroscopy) of the $Mo_{1-x}W_xS_2$ alloys. Insets in a) show the peak positions of the Raman peaks as a function of the supercycle length. Insets in b) show the spectral positions of the extinction peaks A and B, and the similarity of the spectrum of the longest-supercycle sample to a superposition of the pure $MoS_2$ and $WS_2$ spectra, indicating the heterostructure-like nature of samples grown using long supercycle lengths. Spectra are offset vertically for legibility.*

**Impact of atomic ordering on electronic structure**
The effect of atomic ordering on the electronic structure of the $Mo_{1-x}W_xS_2$ films was investigated by optical transmission spectroscopy. Samples were grown using supercycle lengths of 2, 6, 10, 16, 24 and 48 cycles on transparent quartz substrates. The cycle fraction $n/(n+m)$ was again kept fixed at 0.5 (resulting in alloy composition $x = 0.6$), and the total number of ALD cycles was 48 for each sample. The extinction spectra (figure 4b) show two peaks around 1.85 eV and 2.1 eV, labeled A and B. The spectrum of the alloy grown with the longest supercycle (48 cycles) is indistinguishable from a superposition of the spectra of pure $MoS_2$ and $WS_2$ (see inset figure 4a). This is expected since for very long supercycles, the deposited film resembles a heterostructure of $MoS_2$ and $WS_2$, such that the extinction spectrum contains 4 peaks (A and B of $MoS_2$ and $WS_2$). Due to their spectral broadness, not all of these 4 peaks can be resolved, and in the following we deconvolve the extinction spectra using a two-peak model. We will focus our attention on the interpretation of the alloy-like samples made using short supercycles (i.e. up to 10 cycles) and not on the heterostructure-like samples made using longer supercycles, for which the interpretation of a two-peak deconvolution is not straightforward.

The B-peak is seen to shift to higher energy as the supercycle length is shortened (i.e. towards random atomic ordering), while the A-peak shows no significant shift. Between supercycle lengths of 2 and 10 cycles, the B-peak has shifted by approximately 20 meV. Such behavior may be related to small changes in the alloy composition (Mo/W ratio) as a function of the supercycle length[25], or instead the atomic ordering could have a direct effect on the electronic structure of the alloy. To explore the latter option, density functional theory (DFT) electronic structure calculations were carried out. For these calculations, the atomic ordering is quantified using the order parameter $J$:[31]

$$J = \frac{P_{sample}}{P_{random}} = \frac{P_{sample}}{x^2 + (1-x)^2} \quad (4)$$

where $P_{sample}$ is the fraction of neighboring identical metal atoms (Mo-Mo or W-W) of the sample and $P_{random}$ is the fraction of neighboring identical metal atoms when the alloy is randomly mixed. The value of $J$ describes the atomic ordering of the alloy: $J = 1$ describes random mixing, $J > 1$ describes clustering of Mo and W (where larger $J$ means larger clusters) and $J < 1$ describes "anti-clustering", i.e. the preferential neighboring of non-identical metal atoms (Mo-W), tending towards a checkerboard pattern.

Typical values of $J$ for $Mo_{1-x}W_xS_2$ alloys synthesized at high temperatures are around 1.0, i.e. these alloys exhibit random mixing of the transition metal atoms.[31] On the other hand, the value of $J$ for our ALD-grown $Mo_{1-x}W_xS_2$ using a supercycle length of 10 cycles can be estimated from STEM (Figure 3) to be approximately 1.8. However, since the unit cell size needed to perform DFT calculations for such large clusters would lead to prohibitively high computational cost, we limit our calculations to a supercell of 5x5x1 unit cells, such that a maximum value of $J$ of 1.41 can be simulated. The calculated electronic transition energies A and B as a function of the order parameter J are shown in figure 5.

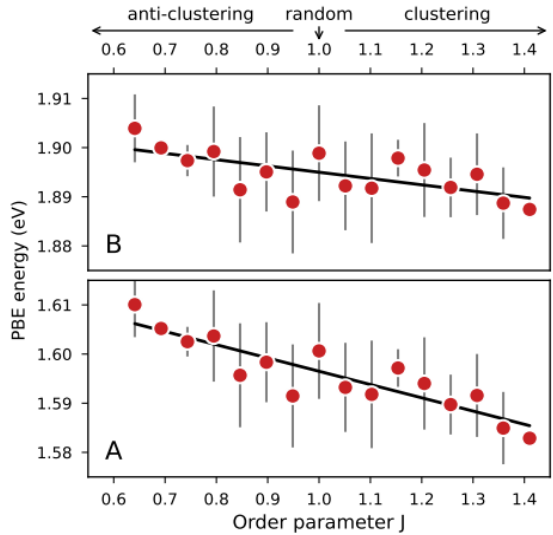

*Figure 5: Calculated fundamental electronic transition energies corresponding to the A (bottom) and B (top) excitons of $Mo_{1-x}W_xS_2$ as a function of atomic ordering. Plotted datapoints are averaged values over 5 unique configurations per order parameter, and the error bars denote their standard deviation.*

Since anti-clustering is not observed experimentally, the results for $J \geq 1$ are most relevant for comparison to experiment. Between $J = 1$ (random mixing) and $J =$ 1.41 (largest clustering), the calculated A transition energy redshifts by 11 meV, whereas the calculated B transition energy redshifts by 5 meV. Linear extrapolation to the appropriate $J$-value of 1.8 for $Mo_{1-x}W_xS_2$ alloys deposited using a supercycle length of 10 cycles yields redshifts of 22 meV and 10 meV for the A and B transitions respectively. These values are of the same magnitude as the experimentally observed shift in the optical extinction spectrum, although the calculations predict a larger shift of the A peak than of the B peak, which was not observed experimentally. Hence, our calculations indicate that atomic ordering indeed could impact the electronic structure of $Mo_{1-x}W_xS_2$, though further study is required to experimentally distinguish the effects of atomic ordering on the electronic structure from the effects of small changes in alloy composition.

## CONCLUSIONS

We have shown that the 2D TMD alloy $Mo_{1-x}W_xS_2$ can be grown with excellent composition control using a supercycle-based atomic layer deposition process. This well-behaved alloy growth allowed us to control the ordering of the transition metal atoms within the individual molecular layers by tuning the supercycle length of the deposition process. This control over atomic ordering was confirmed directly through HAADF-STEM imaging, and the atomic ordering was shown to tune the vibrational spectrum and electronic structure of the alloys. Our results indicate that atomic ordering of 2D semiconductors alloys can be experimentally manipulated during synthesis to finely tune the opto-electronic properties of these materials for specific applications. Furthermore, our results indicate that Raman spectroscopy is sensitive to changes in atomic ordering of 2D TMD alloys, making it an accessible method of probing the atomic ordering of these materials.

## METHODS

### PE-ALD process

Plasma-enhanced ALD of $Mo_{1-x}W_xS_2$ films was performed using an Oxford Instruments FlexAL-2 reactor equipped with a remote inductively coupled plasma (ICP) source. Silicon wafers with 450 nm thermally grown oxide were used as substrates. The substrate table heater was kept at 450 °C during processing, resulting in a substrate temperature of 350 °C; a 20 minutes wait step was performed prior to deposition to ensure thermal equilibration. The ALD processes used for the $MoS_2$ and $WS_2$ cycles are described in detail in previous work.[11,34] As precursors, $Mo(N^tBu)_2(NMe_2)_2$ (98%, Sigma Aldrich) and $W(N^tBu)_2(NMe_2)_2$ (99%, Strem Chemicals) were used for $MoS_2$ and $WS_2$ respectively. The precursors were kept in stainless steel canisters which were heated to 50 °C, and precursor delivery into the reaction chamber was facilitated with a 50 sccm argon bubbling flow. Precursors were dosed for 6 seconds, followed by a 10-second purge of the reactor chamber with 100 sccm argon. A plasma of 10 sccm $H_2S$ and 40 sccm Ar was used as the coreactant in both the $MoS_2$ and the $WS_2$ processes. Plasma exposure was performed at a power of 100 W for 30 seconds at a pressure of approximately 6 mTorr. A subsequent 10-second purge of the reactor chamber with 100 sccm argon completes the ALD cycle.

**Film thickness measurement**
Film growth was monitored by spectroscopic ellipsometry (SE) using a J.A. Woollam M-2000 ellipsometer in the spectral range from 1.25 to 4 eV. The film thickness and optical constants were determined by parametrizing the SE data using a B-spline.[35] Comparison of the optical constants obtained for pure $MoS_2$ to literature values[36] showed good agreement, supporting the validity of the B-spline method for determining the thickness and optical constants of thin (< 10 monolayers) TMD films.

**Elemental composition measurement**
The relative elemental composition of the $Mo_{1-x}W_xS_2$ films was determined through x-ray photoelectron spectroscopy (XPS) using a Thermo Scientific K-alpha spectrometer with an aluminum K-α (1486.6 eV) radiation source. In the experiments on atomic ordering, the obtained atomic abundances were corrected for the exponential attenuation of the emitted photoelectrons with depth. Rutherford backscattering spectroscopy (RBS) was used to determine the absolute elemental composition of selected samples as additional validation. RBS measurements were performed with a 2 MeV $^4$He beam and a detector at 170° scattering angle.

**Raman and PL measurements**
Raman scattering spectroscopy and photoluminescence (PL) spectroscopy measurements were performed with a Renishaw InVia Raman microscope equipped with a 514.5 nm laser and a CCD detector. Raman peak positions corresponding to the $A_{1g}$ modes were extracted by deconvolution of the spectra using a five-peak Voigt model, which is a common strategy used in the literature.[12]

**Optical absorption measurements**
Optical absorption spectroscopy was performed using a J.A. Woollam M-2000D spectroscopic ellipsometer in transmission mode.

**Atomic ordering stochastic simulation**
The simulation starts with an empty hexagonal grid. In a loop, a random grid point is chosen. If this grid point is empty, there is a chance of 0.001 that a "$MoS_2$ unit cell" is placed there (simulating nucleation). If the empty grid point is next to a grid point that is already filled, this chance is 1 (simulating preferential edge growth). This loop is repeated N times (simulating a full ALD cycle). By alternating such virtual ALD cycles of $MoS_2$ and $WS_2$, the full supercycle process of $Mo_{1-x}W_xS_2$ alloy deposition is simulated.

**Ab-initio electronic structure calculations**
DFT calculations were carried out with the projector-augmented-wave[37] (PAW) framework as implemented in the VASP software[38] with the exchange and correlation contributions to the electronic energy described semi-locally by the PBE functionals.[39] Van der Waals interactions were modeled using the DFT-D3 method of Grimme.[40] Supercells of 5x5x1 primitive cells were used in the calculations, which were geometrically optimized down to a residual force tolerance of $10^{-4}$ eV/Å using a converged plane-wave cut-off energy (400 eV) and k-space sampling (3x3x1 Monkhorst-Pack mesh). Effective electronic band structures were obtained through the unfolding procedure.[41] For each degree of atomic ordering, five unique supercells were generated and the obtained electronic transition energies were averaged over these structures.

**TEM studies**
The monolayer films were deposited on ultra-thin (approximately 4.5 nm) $Si_3N_4$ TEM windows. Top view TEM studies were performed using a probe-corrected JEOL ARM 200F, operated at 200 kV. Focusing was performed outside the area of interest. The imaged areas were only exposed to the electron beam in the single scan required of acquiring the image, in order to minimize beam damage effects. For image post-processing, a highpass filter with a cutoff length of 100 pixels was applied to the micrographs to suppress background and allow for lossless contrast enhancement.

**Electronic structure calculations**
Electronic band structure calculations were performed on 5x5x1 supercells. The electronic band-to-band transition energies were derived from the calculated conduction band minimum and valence band maximum at the K-point of the Brillouin zone (where the A and B excitons are located). The band structures were calculated at the PBE level, such that a systematic offset is expected between the calculated transition energies and the experimentally measured absorption peak positions.

**CONFLICT OF INTEREST**
There are no conflicts of interest to declare.

**DATA AVAILABILITY**
The data that support the findings of this study are available from the corresponding author upon reasonable request.

**ACKNOWLEDGEMENTS**
This work was partially funded by The Netherlands Organization for Scientific Research (NWO) through the Gravitation program "Research Centre for Integrated Nanophotonics". This work has also been supported by the European Research Council (Grant Agreement No. 648787-ALD of 2DTMDs). The authors acknowledge C.O. van Bommel, M.G. Dijstelbloem, C.A.A. van Helvoirt, B. Krishnamoorthy and J.J.A. Zeebregts for their technical assistance, and thank S. ter Huurne for performing UV-vis absorption measurements. Solliance and the Dutch province of Noord-Brabant are acknowledged for funding the TEM facility.

Integrated Nanophotonics". This work has also been supported by the European Research Council (Grant Agreement No. 648787-ALD of 2DTMDs). The authors acknowledge C.O. van Bommel, M.G. Dijstelbloem, C.A.A. van Helvoirt, B. Krishnamoorthy and J.J.A. Zeebregts for their technical assistance, and thank S. ter Huurne for performing UV-vis absorption measurements. Solliance and the Dutch province of Noord-Brabant are acknowledged for funding the TEM facility.